\documentclass[11pt]{article}

\usepackage{fullpage}
\usepackage{color}
\usepackage{longtable}
\usepackage{lscape}
\usepackage{todonotes}

\title{Content Confidentiality in Named Data Networking\thanks{This research has been funded by the Defense Advanced Research Projects Agency (DARPA) under contract HR0011-17-C-0111. The views,
opinions, and/or findings expressed are those of the author(s) and should not be interpreted as representing the
official views or policies of the Department of Defense or the U.S. Government.}}
\author{Aleksandr Lenin, Peeter Laud\\ Cybernetica AS\\ \texttt{aleksandr.lenin|peeter.laud@cyber.ee}}

\begin{document}
\maketitle

\begin{abstract}
In this paper we present the design of name based access control scheme which facilitates data confidentiality by applying end-to-end encryption to data published on NDN with flexible fine-grained access control, which allows to define an enforce access policies on published data. The scheme is based on ciphertext-policy attribute-based encryption (CP-ABE). We discuss the use of the scheme on the basis of two use-cases, and report overhead associated with it, based on our implementation.
\end{abstract}

\section{Introduction}

\subsection{Named Data Networking}

Named Data Networking (NDN) is one of five projects funded by the U.S. National Science Foundation under its Future Internet Architecture Program~\cite{zhang_ndn_2014}. Named Data Networking builds upon the results of an earlier project, Content--Centric Networking (CCN), while both NDN and CCN are instances of more general network research direction called Information--Centric Networking (ICN), under which different future network architecture designs have emerged~\cite{Xylomenos_2014,zhang_ndn_2014}. CCN originally proposed transition from today's connection--oriented IP networks to content--centric architectures, and NDN develops these ideas even further. Currently, named data networking exists as a testbed prototype built as overlay network on top of the IP network, but active research is going on to investigate possibilities to replace the IP stack with NDN stack~\cite{ContiGHLL20}. 

The main design principle of NDN and the distinction from the IP networks is that IP networks are communication networks, where packets address nodes, that are communication endpoints in the network, while NDN is a distribution network, which focus on solving the content distribution problem. Solving this problem over communication network is complex and error--prone~\cite{zhang_ndn_2014}.

The main innovation of NDN is the approach to change the transport layer in the network protocol stack, such that packets name content objects, rather than communication endpoints. This changes the semantics of the network from delivering packets to a given destination to fetching data identified by given names~\cite{zhang_ndn_2014}. This change is conjectured to help solving not only end-to-end communications, but also content distribution and control problems. 

There are two types of packets exchanged over NDN --- the interest packets and the data packets. Based on their roles in the network, three types of nodes are distinguished. The producer nodes (or the publisher nodes) are the ones who produce the content in content-centric networks. Consumers are the nodes in the network which request content from the producers. To request content, consumers send an interest packet, which contains a name of the content that the consumer wishes to get. Producers respond with data packets that routed back to the consumer following the same route as the consumer's interest packet. The forwarding nodes, called NDN routers, are the ones responsible for forwarding packets between the producers and the consumers. An NDN router maintains three tables: the Pending Interest Table (PIT), the Forwarding Information Base (FIB), and a Content Store (CS). When an interest packet arrives, forwarding node first checks its content store for possible matches. If the content store contains the requested data, the router serves it back to the consumer. Otherwise, the router searches its PIT for matches. The PIT stores all the interest requests that a forwarding node has forwarded, but has not received any data packets yet. Each PIT entry contains the name of the requested content, together with incoming and outgoing interfaces. PIT records information about interfaces via which the incoming interest requests have arrived and groups them by the content name, thus permitting one content name be related to several incoming interfaces. This allows efficient multicast data delivery. If the matching record in the PIT exists, the router updates the PIT entry by recording the interface where the corresponding interest packet came from. Of no matching records exist, the router forwards the interest packet towards the content producers based on the information contained in its FIB as well as its forwarding strategy, which prescribes when and where to forward the interest. When a router receives a data packet, it forwards it back via all the incoming interfaces listed in the corresponding PIT entry, then removes the PIT entry and caches the data packet in its content store. Routers forward packets based on the names of the content requested or contained in these packets. NDN packets have no information about hosts and their interfaces. Since every data packet is signed by the producer which generated it, the contents of a data packet are meaningful independently of who delivers the content to the consumer. This permits routers to serve the content from their cache to satisfy the interest requests faster due to the delay between request and response.

The NDN design assumes names that belong to a hierarchical namespace which will never suffer from address space exhaustion, since the namespace is unbounded. The naming scheme is designed to be independent of the network thus permitting it to evolve independently. Flat naming schemes are also supported, although structured naming schemes supersede the flat ones~\cite{zhang_ndn_2014}. 

\subsection{NDN security}

NDN follows data-centric security approach, in which the content producer is required to sign all the data packets it generates. This ensures the integrity and authenticity of a data packet. It allows to decouple the consumer's trust from the network node that served the content, and replaces it with the trust towards the producer directly, thus allowing the consumers to decide if they trust a given data producer as a reliable source of the requested content. The trust management is the most challenging part of the NDN architecture, and several models, hierarchical as well as the ones based on the web of trust, have been proposed, but still the problem of establishing trust remains an open question~\cite{zhang_ndn_2014}. 

While integrity properties in data-centric networks are easy to state and compare with similar properties in connection-centric networks, the issues with confidentiality properties are less clear-cut. In a connection-centric network, the integrity of a channel means that (i) it is known where the source of the channel is, and (ii) the messages on the channel are not being tampered with. From these properties one may deduce that whenever one receives a data item over a channel with integrity, then the source of that data item will be known and that source intended to send this data item. The same properties are ensured by the cryptographic primitive of digital signatures, and their preserverance is not affected by the manner the data item moves across the network. A confidential channel is one where (i) its sink is known, and (ii) the messages on the channel cannot be observed. Its analogue in a data-centric network is less clear.

In a data-centric network, end-to-end confidentiality for a piece of data would have to be meaningful independently of considering any server or network node that stores this content. It is definitely possible to use the routing mechanisms of NDN to distribute encrypted content, and also to distribute the decryption keys, protecting the latter with public-key encryption of some sort. In a sense, this would simulate the confidentiality properties in connection-centric networks: a publisher has a list of recipients (with their public keys) for its content, and makes the content available only to them.



Publishers face two problems --- key distribution and key management. In order to grant content access to authorized consumers, the publisher needs to deliver each of the consumers their individual decryption key in a secure way. This can be achieved by publishing as many short messages (encryptions of the key) as there are recipients, each of them meaningful only to a single recipient. A better way would somehow separate out the group of recipients, with the publisher not required to keep a comprehensive list of them, but still somehow matching the intent of the publisher.

In this paper we make no assumptions regarding the existence of an established trust model, nor the existence of trusted authorities in NDN network. We look for a solution that would not increase the complexity of the NDN network by assigning trusted certifying authorities, and will be using self-certifying public keys instead. There exist a relatively recent results in cryptography research which reconsider the concept of a public key cryptography --- the identity based encryption (IBE) and attribute-based encryption (ABE). In a typical public key cryptosystem, a user is assigned with one or more keypairs, and every keypair consists of a private and a public key. Identity based encryption changed the common understanding of public key cryptography by allowing the public key of a user to be an arbitrary string (i.e. email, a phone number, state issued identification code, etc). For anyone to encrypt something using IBE, all one needs to know is the public string identity of the recipient. This identity is directly used for encryption, and the recepient's public key is explicitly derived from this public identity. On the other hand, attribute-based encryption (ABE) has taken it even further away from the traditional understanding of public key cryptography, and defines the user's identity not as one single piece of information, not as one single token, but as a set of strings, which are called attributes. ABE allows to encrypt a message for a set of specific attributes, so that only users holding private keys that match this set of attributes, can decrypt the message. In an ABE scheme, the users' keys are issued by some trusted party, usually called the key distribution center (KDC). Each producer may fulfill the role of a KDC by properly verifying the identities of his clients and issuing private keys to authorized customers. This approach replaces the trust model from globally trusted certifying authorities with trust towards the content publisher, and the publisher itself being the trust anchor is a feature naturally occurring in NDN.

Attribute based encryption can be viewed as a secret sharing scheme with general access structure, encoded by a set of ABE attributes. ABE permits fine-grained access control over the attributes and allows to express access policies in the form, semantically equivalent to a Boolean function as a logic statement over the attributes. Depending on the location of the access structure, two versions of ABE may be distinguished: the ciphertext policy attribute based encryption (CP-ABE) and the key policy attribute based encryption (KP-ABE). In CP-ABE, the ciphertext contains the access structure and defines an access policy over a global set of users' attributes. The users' keys contain attributes that specify users' access permissions, and the users whose key attributes match the access structure in a ciphertext, get access to the shared secret, which is usually the symmetric key. In KP-ABE on the contrary, the access policy is located in users' private keys and specifies which ciphertexts this key is permitted to decrypt, while the ciphertexts contain attributes describing the type of content. Only the ciphertexts whose attributes match the access structure of a given key, will be correctly decrypted by a given key.

\subsection{Our contribution}

For a data-centric access control that NDN ideology strives for, we compare CP-ABE solutions vs KP-ABE solutions. In the case of KP-ABE, the access policy is located in the users' keys, which in itself, is not really data centric. The publisher attaches a set of attributes to the content, specifying which keys are able to decrypt it. Changing the access policy necessitates a re-key process during which users are issued with new private keys containing new updated access policy. In the case of CP-ABE, the access policy is contained within the published content, independently of the user's keys. If there is a need to update the access policy, the publisher can re-encrypt the content with a new access policy, and no changes are required to the users' keys. ABE primitive can be used for broadcast encryption, and in this case it is called attribute-based broadcast encryption (ABBE), and depending whether it is key policy or ciphertext policy version, they are called KP-ABBE and CP-ABBE correspodingly. It more in--line with the data--centric access control ideology to allow the access policy to be attached to the content itself, rather than the keys possessed by the users, and thus CP-ABBE fits our problem the best. 

CP-ABBE also seamlessly integrates with the organizational role hierarchy as described on our second scenario. A publisher can encrypt some content to a particular group of users by encrypting the content to the attributes, shared among all the users of the group. I.e., one may encrypt the content using the attributes "senior academic personnel" and "computer science department", and all the users whose keys have both of these attributes, will be granted access to the published content. The publisher doesn't even need to know who the senior academic personnel in the computer science department exactly are --- all one ever needs to know, is that the intended recipients' share the "senior academic personnel" and "computer science department" attributes. Such abstraction mechanisms make CP-ABBE much more powerful alternative compared to IBE solutions, where one needs to explicitly list all the identities of the recipients.  Abstract attributes offered by CP-ABBE bring along some difficulties related to efficient attribute revocation. Several users may share the same abstract attribute, and quite often there is a need to revoke access of any specific user only, without affecting the others who share the same attribute. Therefore we are focusing our attention on CP-ABBE schemes having efficient revocation capabilities. Revocation in CP-ABE schemes comes at a price, and despite active research being conducted to tackle this problem, there are a handful of solutions that offer efficient revocation capabilities. Such CP-ABBE schemes increase the flexibility of access control at a cost of reduced efficiency, as compared to CP-ABBE schemes without revocation capabilities.

In this paper we present a concrete implementation of a CP-ABBE scheme, its integration with NDN, and measurements of the overheads it creates. We present and discuss two case studies where this implementation is useful; our measurements of overheads are tightly related to the usage scenarios in one of the case studies.

\section{State Of The Art}

\subsection{Ciphertext Policy Attribute Based Encryption}

Bethencourt \emph{et al.}~\cite{Bethencourt2007} present a CP-ABE scheme with monotone access structure. The authors call it the monotonic access tree, which is a tree structure where every node is a threshold gate. This structure naturally supports the Boolean semantics via AND and OR type gates, and besides more general $m$-of-$n$ threshold gates. A key can decrypt the content encrypted under such an access structure, if the assignment of attributes to the tree leaves, corresponding to the attributes in the key, satisfy the access tree. The monotone access structure can be seen as a limitation of the scheme, since it doesn't support efficient attribute revocation. To support non-monotone access structures, for every attribute $s$, the authors suggest to introduce another attribute $\hat{s}$ that would correspond to negated $s$. This construction would result is doubling the number of attributes. since in CP-ABBE schemes, the size of the ciphertext tends to grow linearly in the number of attributes, doubling the number of attributes is equal to doubling the size of the ciphertext. If a user's key is identified with $n$ attributes, the key contains $2n+1$ bilinear group elements, and the key generation requires $2(n+1)$ exponentiations. The ciphertext contains an access tree, a blinded message, and $2k+1$ bilinear group elements, where $k$ is the number of leaf nodes in the access tree. The encryption operation requires $2(k+1)$ exponentiations. If an access tree has $u$ leaves and $v$ non--leaf nodes, decryption requires $2u$ pairing computations, and $v$ exponentiations. The authors prove semantic security in the generic bilinear group model. Although the authors did not prove CCA security, they mentioned that it is possible to do, by applying certain techniques to their scheme. The authors state that their system is collusion--resistant, however they do not quantify the collusion resistance property of their system.

Emura \emph{et al.}~\cite{Emura2009} propose a new Ciphertext-Policy Attribute-Based Encryption (CP-ABE) with constant ciphertext length, where the number of pairing computations is also constant. The access structure is a set of attributes, where every attribute may take on one of the three values: positive (attribute is required), negative (attribute is negated and must not be present in the key), wildcard (the access structure is indifferent w.r.t. the presence or absence of a specific attribute in the key). The scheme supports possibility to add new attributes after the setup phase without breaking the security properties. The authors prove CPA security under the DBDH assumption. The key generation requires $3$ exponentiations, and the user's private key contains $2$ group elements. If $W = [W_1,\ldots,W_n]$ is an access structure, then the encryption operation takes $2$ exponentiations, as well as $|W|$ multiplications, and the cryptogram contains an access structure $|W|$ and $3$ group elements. The decryption operation requires $3$ exponentiations, as well as $2$ pairing computations. 

Zhou \emph{et al.}~\cite{Zhou2010}  proposed a new construction of CP-ABE, named Constant-size CP-ABE (CCP-ABE), and a corresponding broadcast version CCP-ABBE having constant ciphertext size. The scheme uses AND gate access policy, where an attribute is labeled as positive, negative, or wildcard (any). The authors prove CPA security under decisional bilinear Diffie-Hellman exponent (K--BDHE) assumption. If there are $k$ attributes in the system, key generation requires $2k+1$ exponentiations, and the size of the key is also $2k+1$. Encryption requires $2$ exponentiations. The size of ciphertext size, apart from the size of the access structure and the encrypted message, contains $2$ additional group elements, which are bounded by $300$ bytes total. Decryption takes $1$ pairing computation.

Yamada \emph{et al.}~\cite{Yamada2014} propose a CP--ABE scheme with non--monotone access structure and constant size of master public key. The private key contains $4k+2$ bilinear group elements, where $k$ is the number of attributes in the key. The key generation procedure requires $8k+3$ exponentiations. The ciphertext contains $3n+2$ group elements, where $n$ is the number of elements used in access policy. Encryption requires $5n+2$ exponentiations and $1$ pairing computation. Decryption requires $3l+m(2k+1)$ pairing computations, where $l$ is the number of non--negated attributes in the key, and $k$ is the number of negated attributes in the key, such that $k = l+m$. The authors prove selective security under a new n--(B) assumption which is secure in the generic group model.

Li \emph{et al.}~\cite{Li2017} suggest a CP--ABE scheme with non--monotone access structure based on ordered binary recision diagrams (OBDD). Therefore, access policies expressed in the form of OBDDs are semantically equivalent to Boolean functions. The size and generation time of the users' private keys is constant. The private key contains $2$ bilinear group elements, and key generation requires $2$ exponentiations. The ciphertext contains $2+k$ bilinear group elements, where $k$ is the number of valid paths in OBDD, and encryption requires $2+k$ exponentiations. The decryption operation requires OBDD traversal, hence the time complexity is thus proportional to the OBDD size, and it appears that it requires $2$ pairing computations for every branch in OBDD access structure. The authors proved collusion--resistance and CPA under the bilinear DH assumption.

Roy \emph{et al.}~\cite{Roy_securedata} present a CP-ABE scheme that uses access structure in the form of an access tree, and hence supports positive as well as negative attributes, and is semantically equivalent to a Boolean function. One of the distinct features of this encryption scheme is that attributes are allowed to change over time. The user's private key contains $2+4n$ bilinear group elements, and its generation takes $2+4n$ exponentiations, where $n$ is the number of attributes in the key. The ciphertext contains $2+2k+3m$ bilinear group elements, encryption takes $2+2k+3m$ exponentiations, where $k$ is the number of non-negated attributes, and $m$ is the number of negated attributes in an access structure. Decryption requires $2s$ pairing computations, as well as $3t + 1$ exponentiations, where $s$ is the number of non-negated attribute in the leaves of the access tree, and $t$ is the number of non-leaf nodes in the access tree structure.

Lubicz \emph{et al.}~\cite{Lubicz2008} propose a CP-ABBE scheme using monotone AND-gate as an access structure, with revocation capabilities that are external to the access structure. The private key contains $2+n$ bilinear group elements, where $n$ is the number of attributes in the key. Key generation requires $2+n$ elliptic group multiplications. Ciphertext contains $k+3$ bilinear group elements, where $k$ is the number of revoked attributes. Encryption requires $3+k$ elliptic group multiplications, and decryption requires $3$ pairing computations. The authors prove full collusion security in the generic model of groups with pairing.

Li \emph{et al.}~\cite{LiZ15} propose a CP-ABBE scheme that supports any monotone access structure. It can be adapted to support user revocation by doubling the number of attributes and introducing a negated version for every attribute. The authors prove, through a dual system encryption methodology, that their system is fully-secure under the composite order bilinear groups assumption. If $n$ is the number of attributes in the key, $m$ is one less than the total number of users in the system, and $k$ is the number of attributes in an access structure, then the key generation requires $3+n+m$ exponentiations, encryption requires $1$ pairing computation, as well as $3(k+1)$ exponentiations, decryption requires $1$ pairing computations. A private key contains $2+n+m$ bilinear group elements, and a ciphertext contains $3+2k$ bilinear group elements.

Zhang \emph{et al.}~\cite{ZhangY18} propose a CP-ABBE scheme offering recipient anonymity by hiding the access structure. Key generation requires $4+n$ exponentiations, where $n$ is one less than the total number of users in the system. Private key contains $3+n$ bilinear group elements. Encryption takes $1$ pairing computation and $4$ exponentiations. Decryption requires $4$ pairing computations. Ciphertext contains $4$ bilinear group elements.

Asim \emph{et al.}~\cite{AsimIP11} proposes a CP-ABBE scheme that uses AND/OR access tree structure and supports revocation that is external to the access structure. The authors prove their scheme is secure under DBDH assumption. Private key contains $3+n$ bilinear group elements, where $n$ is the number of attributes in the key. Key generation requires $3+n$ exponentiations. Ciphertext contains $3+t+s$ bilinear group elements, where $t$ is the number of attributes in access policy, and $s$ is the number of revoked attributes. Encryption requires $3+t+s$ exponentiations and $2$ pairing computations. Decryption requires $4$ pairing computations and $2$ exponentiations.

Junod \emph{et al.}~\cite{JunodK10} propose a CP-ABBE system with access policies supporting AND,OR,NOT gates. The authors prove semantic security and collusion resistance in generic group model with pairings. Private key contains $2n+s+t$ bilinear group elements, where $n$ is the total amount of attributes in use, $s$ is the number of non-negated attributes in the key, and $t$ is the number of negated attributes in the key. Key generation requires $n$ exponentiations, encryption requires $1$ pairing computation as well as $2+k$ exponentiations, decryption requires $2$ pairing computations as well as $k$ exponentiations. Ciphertext contains $3$ bilinear group elements.

\subsection{NDN confidentiality}

Fotiou \textit{et al.}~\cite{Fotiou2016} mention that traditional forms of encryption introduce significant overhead when it comes to sharing content with large and dynamic groups of users, with proxy re-encryption being a convenient solution. The authors use Identity-Based Proxy Re-Encryption (IBPRE) to provide confidentiality and access control for content items shared over ICN, realizing secure content distribution among dynamic sets of users. The suggested approach does not suffer from key escrow problem, as is the case with a similar approach IB-PRE, and does not require out-of-band secret key distribution.

Mannes \textit{et al.}~\cite{Mannes2015} suggest an access control solution to ICN by adapting and optimizing a proxy re-encryption scheme. The authors state that the proposed solution is perfectly aligned with ICN demands, simultaneously ensuring content protection against unauthorized access of contents retrieved from unrestricted in-network caches as well as access control policies enforcement for legitimate users.

Wood \textit{et al.}~\cite{Wood2014} present a secure content distribution architecture for CCN that is based on proxy re-encryption.  The  design  provides  strong  end-to-end  content security  and  reduces  the  number  of  protocol  messages  required for user authentication and key retrieval. Unlike widely-deployed solutions, the suggested solution is also capable of utilizing the opportunistic in-network caches in CCN. We also experimentally compare two proxy re-encryption schemes that can be used to implement the architecture,  and  describe  the  proof  of  concept  application  the authors developed  over  CCNx.

Misra \textit{et al.}~\cite{Misra2013} propose a novel secure content delivery framework, for an information-centric network, which will enable content providers (e.g., Netflix and Youtube) to securely disseminate their content to legitimate users via content distribution networks (CDNs) and Internet service providers (ISPs).  Use of the framework will enable legitimate users to receive/consume encrypted content cached at a nearby router (CDN or ISP), even when the providers are offline. Our framework would slash system-downtime due to server outages, such as that recently experienced by Netflix, Pinterest, and Instagram users in the US (October 22, 2012). It will also help the providers utilize in-network caches for shaping content transmission and reducing delivery latency. The authors discuss the handling of security, access control, and system dynamics challenges and demonstrate the practicality of the framework by implementing it on a CCNx testbed.

AccConF~\cite{Misra2016} is an efficient access control framework for ICN, which allows legitimate users to access and use the cached content directly, and does not require verification/authentication by an online provider authentication server or the content serving router.

Zhang \textit{et al.}~\cite{Zhang2017} present the design of name based access control (NAC) which provides automated key management by developing systematic naming conventions for both data and cryptographic keys. The authors also discuss an enhanced version of NAC that leverages attribute based encryption mechanisms (NAC-ABE) to improve the flexibility of data access control.

Yu \textit{et al.}~\cite{Yu2016} present the design of Name-based Access Control (NAC), which implements the content-based access control model in Named Data Networking (NDN). The paper demonstrates how to make use of naming convention to explicitly convey access control policy and efficiently distribute access control keys, thus enabling effective access control. The authors evaluate the scalability of NAC against CCN-AC, another encryption-based access control scheme.

Yingdi Yu in his Ph.D thesis~\cite{Yu2016a} introduces a data-centric security model for NDN which consists of two parts: data-centric authenticity and data-centric confidentiality. NDN achieves data-centric authenticity by mandating per packet signature, and data-centric confidentiality by data encryption. The dissertation presents a security framework to automate data-centric security of NDN and reduce the enabling overhead. To achieve that, the author designed NDN certificate system to facilitate public key distribution in NDN; a Trust Schema -- a name-based policy language to specify trust model, in order to automate fine-grained data authentication; a timestamp service De-Lorean to address the authenticity problem of archival data; an access control protocol Name-based Access Control to automate data-centric confidentiality at fine granularities.

Ghali \textit{et al.}~\cite{Ghali2015} state that caching makes it difficult to enforce access control policies on sensitive content, since routers only use interest information for forwarding decisions. The authors introduce Interest-Based Access Control (IBAC) which is a scheme for access control enforcement using only information contained in interest messages. It makes sensitive content names unpredictable to unauthorized parties. It supports both hash- and encryption-based name obfuscation. The solution also addresses interest replay attacks by formulating a mutual trust framework between producers and consumers that enables routers to perform authorization checks before satisfying interests from local caches. The proposed design is flexible and allows producers to arbitrarily specify and enforce any type of content access control, without having to deal with content encryption and key distribution.

Fotiou \textit{et al.}~\cite{Fotiou2012} propose an access control enforcement delegation scheme which enables the purveyor of an information item to evaluate a request against an access control policy, without having access to the requestor credentials nor to the actual definition of the policy. The authors state that such an approach has multiple merits: it enables the interoperability of various stakeholders, it protects user identity and it can set the basis for a privacy preserving mechanism. An implementation of the scheme supports its feasibility.

Hemanathan \textit{et al.}~\cite{Hemanathan2015} introduce a Role Based Content Access Control mechanism which provides the contents specific to user based on the role to which it was assigned. Each and every user is authenticated with an AAA server specifically designed for NDN and is validated against the access control policy. Only if a user has access to the content then the Content Packets will be sent or else access will be denied. In this method, when Content Provider receives Interest Packet from the user, it will be forwarded to AAA server and based on the response, the decision is made. In addition to that, NDN routers will also have an access table which will maintain the content name and the allowable \& deniable enroll ID’s. Based on the access table, it allows or denies the access to content packet or to the content provider. If there is no entry in the access table for the enroll ID, then it adds an entry into Pending Validation Table and sends validation request to Content Provider (CP) which will validate with AAA server and reply back with allow or denial message. 

Kurihara \textit{et al.}~\cite{Kurihara2015} propose a comprehensive encryption-based access control framework for content centric networking (CCN), called CCN-AC. This framework is both flexible and extensible, enabling the specification, implementation, and enforcement of a variety of access control policies for sensitive content in the network. The design of CCN-AC heavily relies on the concept of secure content object manifests and leverages them to decouple encrypted content from access policy and specifications for minimum communication overhead and maximum utilization of in-network caches. To demonstrate the flexibility of framework, the authors also describe how to implement two sample access control schemes, group-based access control and broadcast access control, within CCN-AC framework.

Hamdane \textit{et al.}~\cite{Hamdane2013} use the generic and conceptual access control scheme called UCONABC to propose an optimum and secured data centric access control model. In the proposal, data is protected by encryption and lock password, and the access is managed by a centralized access control list (ACL).

Hamdane \textit{et al.}~\cite{Hamdane2015} propose an encryption based access control solution that does not require prior knowledge of all authorized entities. The solution assigns access rights based on certified encrypted credentials provided by the different entities. A formal security analysis is provided as well.

Shang \textit{et al.}~\cite{Shang2015} present NDN-ACE, a lightweight access control protocol for constrained environments over Named Data Networking (NDN). NDN-ACE uses symmetric cryptography to authenticate the actuation commands on the constrained devices but offloads the key distribution and management tasks to a more powerful trusted third party. It utilizes hierarchical NDN names to express fine-grained access control policies that bind the identity of the command senders to the services they are authorized to access. The key management protocol in NDN-ACE allows the senders to update their access keys periodically without requiring tight synchronization among the devices. The evaluation shows that NDN-ACE has fewer message exchange and uses fewer components in the overall network architecture compared to the IP-based alternatives.

Abdallah \textit{et al.}~\cite{Abdallah2016} propose a Decentralized Access Control Protocol for ICN architectures (DACPI). In this protocol, fewer public messages are needed for access control enforcement between ICN subscribers and ICN nodes than the existing access control protocols. DACPI depends on ICN self-certifying naming scheme. The authors perform security analysis on DACPI for the following attacks: man-in-the-middle, forward security, replay attacks, integrity, and privacy violations. According to the security analysis, DACPI prevents unauthorized access to ICN contents with fewer messages passed.

Zhang \textit{et al.}~\cite{Zhang2011} propose a new name-based trust and security protection mechanism. The scheme is built with identity-based cryptography (IBC), where the identity of a user or device can act as a public key string. In a named content network, a content name or its prefixes can be used as public identities, with which content integrity and authenticity can be achieved using IBC algorithms. The trust of a content is seamlessly integrated with the verification of content's integrity and authenticity with its name or prefix, instead of public key certificate of its publisher. Flexible confidentiality protection is enabled between content publishers and consumers. For scalable deployment purpose the authors propose a hybrid scheme combined with PKI and IBC.

Hamdane \textit{et al.}~\cite{Hamdane2012} propose to enhance security in CCN/NDN projects. The authors first define requirements for their naming system in order to provide security services that bind both naming and content. Then, they propose a hybrid scheme which combines public-key infrastructure (PKI) and Hierarchical Identity-Based Cryptography (HIBC) in order to meet the defined requirements. This proposal represents a defense against a potential attack and perfectly fits in with the structures of the various objects of CCN/NDN.

Kurihara \textit{et al.}~\cite{Kurihara2016} argue that explicitly-given names of content makes the censorship easily enforceable in CCNs. The paper introduces an anonymization framework to circumvent the censorship under the novel concept of consumer-driven access control to interest names and opportunities of cache recycling at network nodes. The framework leverages an arbitrary type of encryption-based access control and enables to recycle the CCN-specific content cache at intermediate nodes in path of the anonymized communication. Furthermore, by combining CCNx manifests and nameless objects with anonymization framework, it becomes possible to maximize the benefit of CCN-specific in-network caching simultaneously with minimizing the computational overhead and circumventing the censorship. The authors claim this is the first anonymization framework for censorship circumvention, which is designed by the CCN-specific approach.

Yu \textit{et al.}~\cite{Yu2015} explores the ability of NDN to enable automated decision making about which key can sign which data and the procedure of signature verification through the use of trust schemas. Trust schemas can provide data consumers an automatic way to discover which keys to use to authenticate individual data packets, and provide data producers an automatic decision process about which keys to use to sign data packets and, if keys are missing, how to create keys while ensuring that they are used only within a narrowly defined scope (“the least privilege principle”). The authors have developed a set of trust schemas for several prototype NDN applications with different trust models of varying complexity. 

Yu \textit{et al.}~\cite{Yu2014} propose an endorsement-based key management system, which is inspired by the concept of Web-of-Trust, to secure ChronoChat, a serverless group chat application over NDN. With the endorsement-based key management system, users in a chatroom can collaboratively authenticate each other’s membership in the chatroom. The system also leverages the synchronization mechanism provided in ChronoChat for efficient key/endorsement distribution and revocation. We further extend the key management system for user identity authentication in a chatroom to enable one user to authenticate another user’s identity without resorting to any external public key infrastructure.

Yu, Y.~\cite{Yu2015a} proposed the new NDN certificate format, discussed several approaches for serving certificates in NDN, and discusses the process of certificate revocation considering the new certificate design.

\section{Case studies}

We consider two case studies, where the set of communicating parties is relatively fixed, changes infrequently, or there are no performance requirements regarding the re-keying operation. 

The first example is of a university that uses a cloud as a storage service. There is a need for fine-grained access control to ensure that content access is granted on a per-need basis, to those who need it for official purposes. Universities have a well established hierarchical structure of the employees' roles, as well as the structural units, and this structure stays fixed throughout a considerable amount of time (usually, several years, until the next major university restructuring). Such a well established hierarchical structure implies that in the corresponding secret sharing scheme, the access structure stays more or less fixed, and even if the solution implies a major re-key operation whenever the access structure changes, this might be acceptable, due to the scarcity of events which result in any changes to the access structure. Sometimes it may be needed to revoke a particular user's access to a published material, thus a revocation mechanism must to be in place. Revoked users should not be able to get access to any content that was published after the revocation event, even if several revoked users collude. For pre-existing content, such a revocation requires re-encryption of the content with a different access structure, in which a particular user or a group of users  is explicitly marked as revoked. This is not the only use of the revocation mechanism, however. It is handy to exploit the hierarchical structure of the organizational units, and to assign specific attributes to those units. Quite often, information needs to be shared to the entire bigger organizational unit, with the exception for some individual subgroups. In this case, it is very efficient still to encrypt the content for the attribute of a bigger organizational unit, and to revoke access for specific smaller sub-units to enforce the required access policy.

The second example is of a group discussions. Consider a fixed set of users who wish to communicate in such a way that any participant can dynamically create a chat room with any subset of the set of all the other participants. The list of all the participants, as well as their credentials, is known to everyone. One participant may be a part of several chat rooms. User revocation is not much of an issue in the second example, assuming that there is no requirement to preserve the conversation history. New chat rooms can be created dynamically as needed. Every time a user joins or leaves a chat room, a re-key operation is performed, and the remaining group participants get their new credentials to continue communication. Technically, this is the creation of a new chat room with increased or reduced number of participants.

Both examples describe use cases for a secret sharing scheme with general (non-monotone) access structures. The reasons behind this choice is that uniform access structures (as opposed to general ones) lack expressivity and do not provide means for fine-grained access control. The non-monotonicity assumption widens the range of possible applications we wish to tackle and facilitates greater flexibility w.r.t. possible set of solutions. It can be seen that in the second example described above, the monotone access structure will not work for the university case, since we might wish to share the contents with all members of a research group, but we want to exclude some certain individuals, which would break monotonicity.

\section{Performance Measurements}


\subsection{Setup}

To begin with, we describe our hardware and software platforms we used to run the benchmarking experiments, as well as describe the prototype implementation of the Lubicz-Sirvent CP-ABBE cryptosystem.

\subsubsection{Hardware Setup}

The experiments were run on a Lenovo X1 Carbon notebook having the following specifications:
\begin{itemize}
\item CPU: Intel Core i7-5600U CPU @ 2.60GHz
\item CPU cache: 32KB (L1i/d), 256K (L2), 4096MB (L3)
\item CPU cores: 4
\item RAM: 7.7 GiB
\end{itemize}

\subsubsection{NDN network and tools}

The following NDN tools and their corresponding versions have been used for running the experiments:
\begin{itemize}
\item NFD 0.7.0-36-gbc0e617e
\item ndn-tools 0.7-15-g3527558
\end{itemize}

The NDN network contained one single NFD node, with \texttt{ndnputchunks} acting as publisher, and \texttt{ndncatchunks} acting as consumer.

\subsubsection{Software Setup}

The experiments were run by Python 3 interpreter running in Debian Linux 10 (Buster) operating system.
%
%
The Lubicz-Sirvent cryptosystem was implemented in C++ and uses on libpari library for algebraic computations. 

The prototype implementation is a toolset that consists of:
\begin{itemize}
\item the elliptic curve calculator --- a tool that generates pairing-friendly curves from Barreto-Naehrig family with 128-bit security level.
\item the key generation tool --- a tool that, given a json formatted configuration file, generates master public key, master secret key, as well as private keys for the users. The tool dumps the generated keys in json format in a file specified by the user on the command line.
\item the encryption tool --- a tool that, given a json formatted configuration file, json formatted file containing keys, generates a session key and wraps it in ABBE header, which is dumped into a json header file supplied by the user on the command line. The access structure for this encryption is described in the json configuration file.
\item the decryption tool --- a tool that takes the json configuration file, json keys file, json header file, decrypts the ABBE header with the given user's private key and displays the session key, if the user is a valid recipient of the encryption, or notifies the user that he is not an intended recipient otherwise.
\end{itemize}

\subsection{Performance measurement experiments}
In this section, we report on the benchmarking results of the Lubicz-Sirvent CP-ABBE scheme coupled with sharing files over NDN. 

\subsubsection{Experiment 1}

First, we measure the time it takes to generate the master public key (MPK), as well as individual master secret keys (MSK) for varying number of users. The number of users is varied starting from $1$,$2$,$5$,$10$, then from $10$ to $100$, increasing in steps of $10$. Every user is assigned with $3$ attributes, uniformly selected at random from a pool of $50$ attributes. 

The Python script that runs the tests performs the following actions:
\begin{enumerate}
\item Fills in the template of the json formatted configuration file that contains the description of the Barreto--Naehrig elliptic curve having 128--bit security level, and populates this template with descritions of $1,2,5,10$ users, each having $3$ attributes, randomly selected from a pool of $50$ attributes. A separate configuration file is created for every number of users.
\item Runs the key generation tool with each of the configuration files generated in the previous step, and measures the time it takes for the ABBE implementation to complete this task. Only the time of the last step is measured, the time it takes to generate the configuration files is not included in this report.
\end{enumerate}

The results are given in Table~\ref{tbl:mpk_msk_gen_time_fixed_attrs}.

\begin{table}[!h]
\caption{MPK and MSK generation time, fixed number of attributes}
\label{tbl:mpk_msk_gen_time_fixed_attrs}
\centering
\begin{tabular}{cc}
Number of users & Generation time (s)\\\hline
$1$ & $0.24$ \\
$2$ & $0.39$ \\
$5$ & $0.79$ \\
$10$ & $1.31$ \\
$20$ & $1.84$ \\
$30$ & $2.05$ \\
$40$ & $2.26$ \\
$50$ & $2.43$ \\
$100$ & $3.17$ \\
$150$ & $4.2$ \\
$200$ & $5.42$ \\
$250$ & $5.72$ \\
$300$ & $6.45$ \\
$350$ & $7.35$ \\
$400$ & $8.11$ \\
$450$ & $9.31$ \\
$500$ & $10.03$ \\
$550$ & $10.06$ \\
$600$ & $10.68$ \\
$650$ & $12.18$ \\
$700$ & $12.94$ \\
$750$ & $13.0$ \\
$800$ & $14.68$ \\
$850$ & $15.73$ \\
$900$ & $15.29$ \\
$950$ & $16.06$ \\
$1000$ & $16.78$
\end{tabular}
\end{table}

\subsubsection{Experiment 2}

In this experiment, we measure the time it takes to generate MSK and individual MPK keys for the users with varying number of attributes. First, we consider the case of $2$,$5$,$10$,$20$,$30$,$50$ random uniformly selected attributes, and afterwards increase the number of attributes in the range $100$ to $1000$ in steps of $50$. 

Similarly to the previous experiment, the Python script performs the following actions:
\begin{enumerate}
\item Fills in the template of the json formatted configuration file that contains the description of the Barreto-Naehrig elliptic curve having 128-bit security level, and populates this template with the descrition of a single user, each having varying number of attributes. A separate configuration file is created for every number of attributes under consideration.
\item Runs the key generation tool with each of the configuration files generated in the previous step, and measures the time it takes for the ABBE implementation to complete this task. Only the time of the last step is measured, the time it takes to generate the configuration files is not included in this report.
\end{enumerate}

The results are given in Table~\ref{tbl:mpk_msk_gen_time_varying_attrs}.

\begin{table}[!h]
\caption{MSK and MPK generation time, varying number of attributes}
\label{tbl:mpk_msk_gen_time_varying_attrs}
\centering
\begin{tabular}{cc}
Number of attributes & Generation time (s)\\\hline
$2$ & $0.13$ \\
$5$ & $0.21$ \\
$10$ & $0.4$ \\
$20$ & $0.77$ \\
$30$ & $1.09$ \\
$40$ & $1.45$ \\
$50$ & $1.78$ \\
$100$ & $3.49$ \\
$150$ & $5.27$ \\
$200$ & $6.98$ \\
$250$ & $8.82$ \\
$300$ & $11.07$ \\
$350$ & $12.42$ \\
$400$ & $14.11$ \\
$450$ & $15.87$ \\
$500$ & $17.71$ \\
$550$ & $19.59$ \\
$600$ & $22.48$ \\
$650$ & $23.44$ \\
$700$ & $25.39$ \\
$750$ & $27.13$ \\
$800$ & $30.78$ \\
$850$ & $31.09$ \\
$900$ & $34.4$ \\
$950$ & $36.68$ \\
$1000$ & $38.8$
\end{tabular}
\end{table}

\subsubsection{Experiment 3}

In the third measurement experiment, we measure the time it takes for varying number of users to download files of varying sizes over NDN. We consider $1$,$2$,$5$, and $10$ users, as well as files of $50$ MiB, $100$ MiB, and $500$ MiB size. Users download a file in parallel, the time is measured for every user individually, as well as the worst time among all of them. 

To run the experiment, first files of $50$ MiB, $100$ MiB, and $500$ MiB are generated by the Linux \texttt{dd} command using \texttt{/dev/urandom} device as the source.

Prior to the commencement of the experiment, these following files are published in NDN using the \texttt{ndnputchunks} tool from \texttt{ndn--tools}:
\begin{enumerate}
\item \texttt{/data/file50M.bin} --- $50$ MiB file
\item \texttt{/data/file100M.bin} --- $100$ MiB file
\item \texttt{/data/file500M.bin} --- $500$ MiB file
\end{enumerate}

For any of the $1,2,5,10$ groups of users, the Python script runs the same amount of threads, one thread per user. I.e. to simulate the actions of $10$ users, $10$ threads are launched. Each thread retrieves all the published files using \texttt{ndncatchunks} tool for every file under consideration and dumping its contents to the filesystem. We measure the time it takes to download the file and get its contents. The results are given in Table~\ref{tbl_ndn_file_dnld_perf}.

\begin{table}[!h]
\caption{NDN unencrypted file download performance}
\label{tbl_ndn_file_dnld_perf}
\centering
\begin{tabular}{p{1cm}p{1cm}ccccccccccp{1cm}}
File size (MB) & Number of users & $1$ & $2$ & $3$ & $4$ & $5$ & $6$ & $7$ & $8$ & $9$ & $10$ & worst time\\\hline
$50$ & $1$ & $1.11$ & & & & & & & & & & $1.11$\\
$50$ & $2$ & $4.29$ & $0.79$ & & & & & & & & & $4.29$\\
$50$ & $5$ & $1.17$ & $0.79$ & $1.81$ & $2.45$ & $0.78$ & & & & & & $2.45$\\
$50$ & $10$ & $1.26$ & $0.87$ & $3.06$ & $0.95$ & $1.0$ & $0.84$ & $1.06$ & $1.88$ & $1.84$ & $2.23$ & $2.23$\\
$100$ & $1$ & $1.83$ & & & & & & & & & & $1.83$\\
$100$ & $2$ & $1.79$ & $2.77$ & & & & & & & & & $2.77$\\
$100$ & $5$ & $1.76$ & $1.75$ & $5.25$ & $1.32$ & $3.06$ & & & & & & $5.25$\\
$100$ & $10$ & $2.17$ & $1.58$ & $1.91$ & $2.48$ & $3.27$ & $1.32$ & $1.52$ & $3.52$ & $4.23$ & $1.41$ & $4.23$\\
$500$ & $1$ & $11.56$ & & & & & & & & & & $11.56$\\
$500$ & $2$ & $11.98$ & $11.46$ & & & & & & & & & $11.98$\\
$500$ & $5$ & $10.38$ & $10.39$ & $13.32$ & $10.21$ & $12.32$ & & & & & & $13.32$\\
$500$ & $10$ & $10.48$ & $14.33$ & $11.83$ & $15.09$ & $14.04$ & $13.25$ & $14.07$ & $14.22$ & $13.84$ & $13.3$ & $15.09$\\
\end{tabular}
\end{table}

\subsubsection{Experiment 4}

This experiment measures the time it takes for $1,2,5,10$ users to simultaneously download an encrypted files of $50,100,500$ MiB over NDN, decrypt the ABBE header, obtain the symmetric key and decrypt the file.

To run the experiment, first files of $50$ MiB, $100$ MiB, and $500$ MiB are generated by the Linux \texttt{dd} command using \texttt{/dev/urandom} device as the source.

An encryption Python script is then launched which, for every file under consideration, executes the following actions:
\begin{enumerate}
\item Launches CP-ABBE key generation tool to generate a new set of keys
\item Launches CP-ABBE encryption tool to get an ABBE header which is stored as \texttt{header.json} file
\item Extracts the session key from the output of the CP-ABBE encryption tool and encrypts all the files using AES with the session key.
\end{enumerate}
The script produces $3$ encrypted files: \texttt{file50M.aes}, \texttt{file100M.aes}, and \texttt{file500M.aes}.

Prior to the commencement of the experiment, these following files are published in NDN using the \texttt{ndnputchunks} tool from \texttt{ndn-tools}:
\begin{enumerate}
\item \texttt{/headers/header.json} --- the header file produced by CP-ABBE
\item \texttt{/data/file50M.aes} --- AES encrypted $50$ MiB file
\item \texttt{/data/file100M.aes} --- AES encrypted $100$ MiB file
\item \texttt{/data/file500M.aes} --- AES encrypted $500$ MiB file
\end{enumerate}

For any of the $1,2,5,10$ groups of users, the Python script runs the same amount of threads, one thread per user. For each file, each thread performs the following actions:
\begin{enumerate}
\item Uses \texttt{ndncatchunks} tool to retrieve \texttt{header.json} and the encrypted file over NDN
\item Launches CP---ABBE decryption tool to decrypt the ABBE header and obtain the AES key
\item Decrypts the encrypted file with the AES key
\end{enumerate}
We measure the time it takes for every thread to download all the relevant files and to decrypt them. The results are given in Table~\ref{tbl_ndn_enc_file_dnld_perf}.

\begin{table}[!h]
\caption{NDN encrypted file download and decryption performance}
\label{tbl_ndn_enc_file_dnld_perf}
\centering
\begin{tabular}{p{1cm}p{1cm}ccccccccccp{1cm}}
File size (MB) & Number of users & $1$ & $2$ & $3$ & $4$ & $5$ & $6$ & $7$ & $8$ & $9$ & $10$ & worst time\\\hline
$50$ & $1$ & $2.33$ & & & & & & & & & & $2.33$\\
$50$ & $2$ & $4.52$ & $2.13$ & & & & & & & & & $4.52$\\
$50$ & $5$ & $7.64$ & $2.16$ & $2.32$ & $2.02$ & $2.02$ & & & & & & $7.64$\\
$50$ & $10$ & $5.82$ & $2.3$ & $4.38$ & $3.86$ & $2.34$ & $2.06$ & $2.18$ & $2.81$ & $2.15$ & $4.59$ & $5.82$\\
$100$ & $1$ & $3.74$ & & & & & & & & & & $3.74$\\
$100$ & $2$ & $3.37$ & $6.21$ & & & & & & & & & $6.21$\\
$100$ & $5$ & $6.17$ & $4.04$ & $5.26$ & $4.49$ & $3.78$ & & & & & & $5.26$\\
$100$ & $10$ & $3.55$ & $6.66$ & $3.42$ & $6.17$ & $3.66$ & $6.58$ & $5.37$ & $4.28$ & $4.27$ & $3.69$ & $6.66$\\
$500$ & $1$ & $16.29$ & & & & & & & & & & $16.29$\\
$500$ & $2$ & $20.87$ & $22.26$ & & & & & & & & & $22.26$\\
$500$ & $5$ & $25.71$ & $22.97$ & $19.89$ & $22.3$ & $22.06$ & & & & & & $25.71$\\
$500$ & $10$ & $28.36$ & $29.05$ & $25.44$ & $29.01$ & $21.7$ & $22.48$ & $19.14$ & $24.52$ & $18.83$ & $20.81$ & $29.05$\\
\end{tabular}
\end{table}

\section{Applying CP-ABBE to the second case study}

For the second case study we consider using CP-ABBE encryption in a chat running over NDN. CP-ABBE flexible access policies facilitate the creation of virtual chatrooms. No separate chat instances are required if CP-ABBE is handling access permissions. 

Instead of keeping a chat instance per chatroom, it is possible to encrypt the messages to specific set of recipients using CP-ABBE and broadcast AES-encrypted message, together with ABBE-encrypted symmetric key (ABBE header) to everyone. The chat client would match the user's private key against the published ABBE header. If the header's access policy permits the user to decrypt the message, the user is a valid recipient. In this case, the chat application will proceed decrypting the message and delivering it to the user. By inspecting the header's access policy and grouping messages by them, a chat application can maintain a set of virtual chatrooms, where every message, posted in the chatroom, will be encrypted with the same access policy. All other participants of such a chatroom will be able to receive and decrypt the message.

The overheads in this use-case are mostly related to synchronizing the views of different parties in regards of which messages have been posted by which participants of the chat~\cite{ZhuA13}. The use of encryption for short messages should not significantly increase it.

\section{Conclusion}

In this paper, we have demonstrated an application of ciphertext-policy attribute-based encryption taking the role of a flexible access policy moderator in secure group communication over the NDN. We have applied a prototype implementation of the Lubicz-Sirvent CP-ABBE system to one of the case studies, the secure data sharing over NDN and reported on its performance characteristics. The results show that the operations of key generation, encryption and decryption are manageable and CP-ABBE can be used in practice. We have also analyzed the merits of CP-ABBE in the case of dynamic groups in use case 2 and conclude that CP-ABBE can be applied to this use case as well, and would offer merits compared to other alternative conventional approaches.

\bibliographystyle{plain}
\bibliography{bibliography/ndn.bib,bibliography/confidentiality.bib}

\end{document}